\documentclass[a4paper,oneside,twocolumn,notitlepage]{extarticle_ecoc}
\usepackage{color}
\usepackage{graphicx}
\PassOptionsToPackage{normalem}{ulem}
\usepackage{ulem}
\pdfoutput=1

\makeatletter

\pdfpageheight\paperheight
\pdfpagewidth\paperwidth

\usepackage{enumitem}

\usepackage{ecoc}
\usepackage[protrusion=true,expansion=true]{microtype}

\makeatother

\usepackage[style=ieee]{biblatex}
\begin{document}

\title{Tailored Shaping, Improved Detection, Simpler Backpropagation: the
Road to Nonlinearity Mitigation}

\author{Marco Secondini\textsuperscript{(1,2)}, Stella Civelli \textsuperscript{(1,2)},
Enrico Forestieri\textsuperscript{(1,2)},}

\maketitle
\begin{strip}
\begin{author_descr}

\textsuperscript{(1)} Tecip Institute, Scuola Superiore Sant'anna
\textcolor{blue}{\uline{marco.secondini@santannapisa.it}}

\textsuperscript{(2)} PNTLab, Consorzio nazionale interuniversitario
per le telecomunicazioni (CNIT)
\end{author_descr}

\end{strip}\setstretch{1.07}

\renewcommand\footnotemark{} \renewcommand\footnoterule{} \let\thefootnote\relax\footnotetext{978-1-6654-3868-1/21/\$31.00 \textcopyright 2021 IEEE} 

\begin{strip}
\begin{ecoc_abstract}
Several strategies for nonlinearity mitigation based on signal processing
at the transmitter and/or receiver side are analyzed and their effectiveness
is discussed. Improved capacity lower bounds based on their combination
are presented.
\end{ecoc_abstract}
\end{strip}

\section{Introduction}

Study and mitigation of nonlinear effects have been an important aspect
of fiber-optic communication since the very beginning\cite{nl-agrawal}.
Aided by digital signal processing (DSP), coherent detection allows
for optical systems with almost unlimited capability to modulate,
demodulate, and process optical signals, improving spectral efficiency
and pushing optical networks toward a seeming capacity limit due to
nonlinear effects\cite{Ellis10,secondini_JLT2017_scope}. This has
further stimulated the research of DSP techniques to overcome fiber
nonlinearity limitations \cite{Roadmap2016}.

The problem of mitigating nonlinear effects and improving system
performance has been addressed from many different perspectives. In
this work, only digital techniques to be implemented at the transmitter
(TX) or receiver (RX) are considered. Therefore, we do not account
for techniques that modify the fiber link or work at optical level,
such as regeneration or phase conjugation.

Following\cite{agrell2018information}, we consider the system described
in Fig.~\ref{fig:System-description}. Fiber propagation is governed
by the Manakov equation\cite{Wang99} and includes periodic in-line
amplification and the simultaneous propagation of other WDM channels,
all independently modulated and detected and with the same input distribution\cite{Agrell:jlt2015}.
Collecting input and output symbols into the vectors $\mathbf{x}$
and $\mathbf{y}$, respectively, we try to maximize the achievable
information rate (AIR)\cite{agrell2018information} by optimizing
the input distribution $p(\mathbf{x})$ (blue block), the detection
metric $q(\mathbf{y}|\mathbf{x})$ (pink block), and TX/RX processing
(yellow blocks). Moreover, we discuss the main modulation/demodulation
and coding/decoding techniques to implement such an optimized system.

\section{Constellation shaping}

\begin{figure}
\begin{centering}
\includegraphics[width=1\columnwidth]{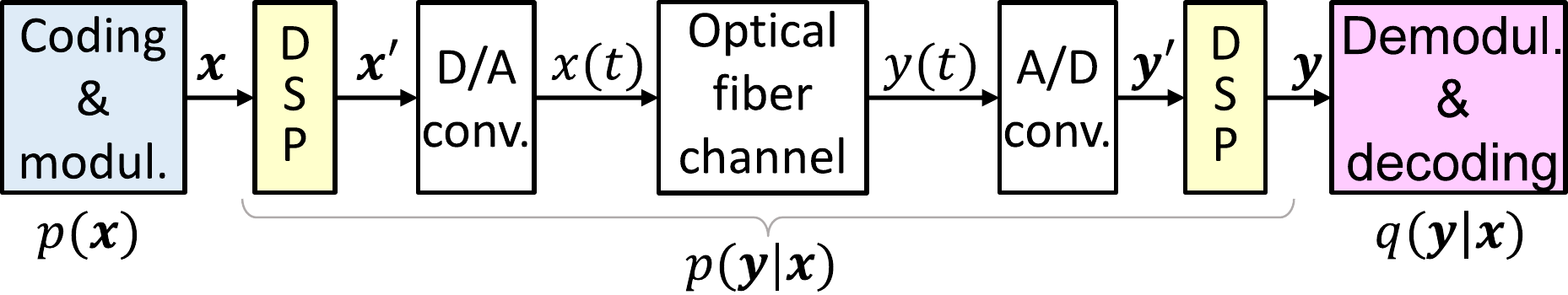}
\par\end{centering}
\caption{\label{fig:System-description}Description of the system considered
for AIR computation and maximization.}
\end{figure}
Constellation shaping improves the efficiency of a digital modulation
scheme by modifying the position of the symbols in the constellation
diagram (geometric shaping) or the frequency with which they are used
(probabilistic shaping). With reference to Fig.~\ref{fig:System-description},
it consists in optimizing the input distribution $p(\mathbf{x})$
(or its support) and devising a proper coded modulation scheme to
encode information accordingly. On the AWGN channel, the problem is
well known: the optimal distribution factorizes into the product of
identical marginal distributions (i.i.d. symbols)---Gaussian in the
general case\cite{shannon48}, Maxwell--Boltzmann (MB) if the symbols
are constrained on a given discrete alphabet\cite{kschischang1993optimal}---which
minimize the energy per symbol required to achieve a certain information
rate. In this context, a practical coded modulation scheme that has
attracted much interest in recent years is probabilistic amplitude
shaping (PAS), thanks to its nearly optimal performance, simple implementation,
and fine rate granularity\cite{bocherer2015bandwidth,buchali2016JLT}.
PAS uses a distribution matcher, followed by a systematic FEC encoder,
to induce the desired distribution over a QAM constellation. The
optimal condition of i.i.d. MB symbols is approached as the block
length of the distribution matcher goes to infinity\cite{bocherer2015bandwidth,gultekin2019TWireless}.

Constellation shaping can be used also to mitigate nonlinear effects.
In this case, often referred to as nonlinear constellation shaping,
the location or probability of the constellation symbols are optimized
to minimize the amount or impact of the generated nonlinear interference
(NLI). There are many evidences suggesting that optimizing the marginal
distribution of i.i.d. 2D symbols yields negligible gains in this
case \cite{fehenberger2016JLT}. In fact, to unlock the full potentiality
of nonlinear constellation shaping, the optimization should be performed
in a higher dimensional space. So far, the approaches have been limited
to the optimization of low-rate constellations in a low-dimensional
space (e.g., geometric shaping in 4D and 8D\cite{kojima2017nonlinearity,chen2019polarization}),
or to a highly constrained optimization of PAS in a higher-dimensional
space (e.g., optimizing the block length of the distribution matcher
\cite{geller2016shaping,fehenberger2020mitigating}). The advantages
obtained in this way are moderate, and might become negligible in
the presence of carrier recovery algorithms \cite{civelli2020interplayECOC}.

The current research challenge is the full optimization of the constellation
in a high-dimensional space, possibly in combination with improved
detection strategies. While this is an extremely complex and still
unsolved problem, in this work we use a recently proposed sequence
selection (SS) bounding technique to estimate the gain achievable
by such an optimization\cite{civelli2021sequence}.

\section{Detection}

Optimal detection requires knowledge of the conditional distribution
$p(\mathbf{y}|\mathbf{x})$. The problem has been widely studied for
the AWGN channel, where $p(\mathbf{y}|\mathbf{x})$ factorizes into
the product of marginal Gaussian distributions, so that optimal detection
can be easily implemented. For the nonlinear fiber channel, $p(\mathbf{y}|\mathbf{x})$
is unknown, so that a mismatched detection based on an approximated
distribution $q(\mathbf{y}|\mathbf{x})$ is used. Often---for simplicity
and in the absence of a suitable alternative---$q(\mathbf{y}|\mathbf{x})$
is still taken as the product of marginal Gaussian distributions,
as in the AWGN channel, but increasing the variance to account also
for NLI.

The search for more accurate and mathematically tractable mismatched
channel models is the subject of current research. For instance,
several models show that interchannel NLI includes relevant phase
and polarization noise (PPN) components that evolve slowly in time
\cite{Sec:PTL2012,Dar2013:opex,dar_JLT2017_nonlinear}. Such components
depend also on frequency and can be alternatively represented as time-varying
linear ISI \cite{secondini2013achievable}. Their mitigation is possible\cite{Sec:PTL14,dar_JLT2017_nonlinear}
and yields an increase of the AIR, which is more effective if combined
with subcarrier multiplexing \cite{Mar:OFC15,secondini2019JLT} and
an optimized per-subcarrier power allocation \cite{garcia2020mismatched1,garciagomez2021mismatched}.
Moreover, even the additive component of NLI has some correlation
in time, which might be exploited for its mitigation \cite{garcia2020mismatched1,garciagomez2021mismatched}.

Another important research topic is the practical implementation of
such improved detection strategies with a reasonable computational
complexity. Besides particle filtering techniques \cite{Dauwels:TrIT08,secondini2019JLT}---useful
for accurate AIR estimation with complex metrics but computationally
too expensive for practical implementation---various approximated
implementations based, e.g., on maximum likelihood sequence detection,
(extended) Kalman filtering/smoothing, recursive least square equalization,
and turboequalization have been proposed \cite{da2018perturbation,golani2019nlin}.
Relevant gains are obtained, but a further reduction of the computational
complexity might be required to make the approach attractive.

\section{Digital backpropagation}

One of the most popular DSP technique for nonlinearity mitigation
is digital backpropagation (DBP), a channel inversion technique that
can be implemented at TX or RX to remove intrachannel NLI\cite{Essiambre:ECOC05,kahn_bp}.
The most classical DBP implementation is based on the split-step Fourier
method (SSFM)\cite{nl-agrawal}, with an accuracy and complexity that
increase with the number of steps. The gains achievable with DBP are
well studied, both numerically and experimentally \cite{savory2010impact,ghazisaeidi2016submarine,dar_JLT2017_nonlinear}.
Unfortunately, reasonable gains require many steps (one or more per
span), so that the search for less complex algorithms is still in
progress. Possible approaches include perturbation methods and Volterra
equalizers \cite{guiomar_opex_2012volterra}; the simplification
of the SSFM\cite{Rafique:OE2011}; machine learning techniques \cite{jarajreh2014artificial,hager_OFC_2018learnedDBP,karanov2018end};
and a combination of the previous approaches---e.g., where a perturbation
method is used to improve the accuracy of the nonlinear step of the
SSFM \cite{secondini_PNET2016}, or machine learning techniques are
used to optimize and simplify a processing scheme inspired by the
SSFM or by a perturbation model\cite{oliari_JLT_2020learnedDBP,redyuk_JLT_2020compensation}.

Another interesting research topic is the possibility to include interchannel
effects in DBP without increasing its complexity, either to improve
the performance by jointly backpropagating several channels, or to
reduce the complexity by means of subband processing in a single channel\cite{mateo2010efficient,hager2018wideband,civelli2021coupled}.

In this work, we simply consider ideal single-channel DBP as an ultimate
limit for intrachannel NLI compensation, and study the AIR gain it
provides when used alone, or in combination with improved shaping
and detection strategies.
\begin{figure*}[!t]
\begin{centering}
\includegraphics[width=1\textwidth]{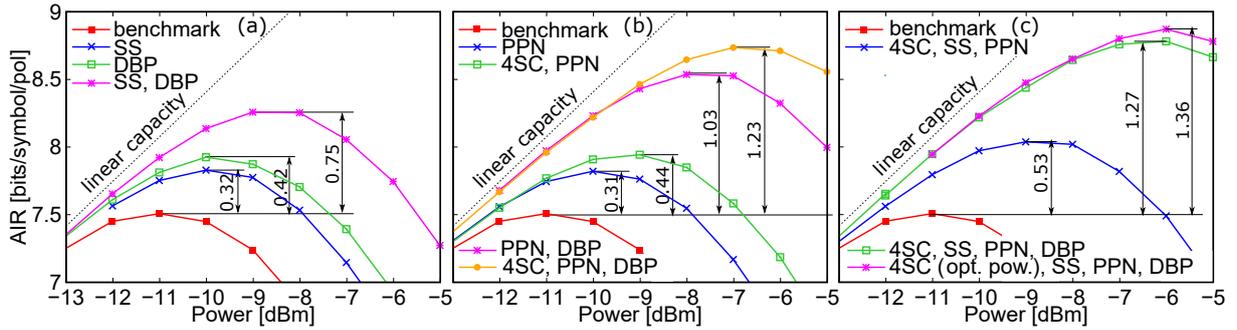}
\par\end{centering}
\caption{\label{fig:AIR}AIR for different combinations of input distribution,
detection metric, and processing.}
\end{figure*}

\section{Achievable information rates}

We investigate and compare the effectiveness of the techniques discussed
above in terms of AIR. The system is depicted in Fig.\ref{fig:System-description},
while the scenario and link parameters are the same considered in
\cite{secondini2019JLT}---a 1000~km standard single-mode fiber
link, ideal distributed amplification, and five 50~GBd Nyquist-WDM
channels with 50~GHz spacing.

As a benchmark, we consider the AIR obtained on the nonlinear channel
when the system is optimized in the absence of nonlinear effects,
i.e., when considering ideal electronic dispersion compensation (EDC),
i.i.d. Gaussian input symbols, and AWGN detection. This benchmark
AIR, reported in Fig.~\ref{fig:System-description}(a)--(c) with
a solid red line, reaches a peak of about 7.5~bits/sym/pol at a launch
power of -11~dBm per channel per polarization, then decreases again.
The other curves are obtained by modifying modulation, detection,
and/or processing with respect to the benchmark, as indicated by the
corresponding labels. The linear capacity $C=\log_{2}(1+\mathrm{SNR})$
is also reported as a reference.

First, we investigate the gains achievable by optimizing the input
distribution. The optimization employs the SS procedure mentioned
above and described in \cite{civelli2021sequence} to minimize the
average variance of intrachannel NLI, with a selection rate of 0.2\%
and a block length of 256 dual-polarization symbols. Besides the benchmark,
Fig.~\ref{fig:System-description}(a) shows three different cases:
SS-optimized input distribution and EDC; i.i.d. Gaussian inputs and
DBP; optimized input combined with DBP. AWGN detection is considered
in all the cases. The SS optimization, alone, yields an AIR gain of
0.32~bits/sym/pol with respect to the benchmark. By comparison, DBP
yields a slightly higher gain of 0.42~bits/sym/pol, while the combination
of the two techniques yields a higher total gain of 0.75~bits/sym/pol.
This suggests that the input distribution provided by SS, though optimized
to reduce intrachannel NLI, partly reduces also interchannel NLI.

Then, we investigate the gains achievable by improving the detection
strategy. Besides the benchmark AIR, Fig.~\ref{fig:AIR}(b) shows
the AIR obtained with PPN detection (with one or four subcarriers,
the latter denoted as 4SC) \cite{secondini2019JLT}, either with EDC
or combined with DBP. In all the cases, i.i.d. Gaussian inputs are
considered. PPN detection works better when combined with subcarrier
modulation \cite{secondini2019JLT}, providing a gain of about 0.44~bits/sym/pol,
comparable with DBP. However, while PPN detection mainly addresses
interchannel NLI, DBP removes only intrachannel NLI. As a result,
when used alone, their effectiveness is limited by the remaining uncompensated
effect. On the other hand, their combination acts synergically, mitigating
both effects and yielding a much higher gain of about 1.23~bits/sym/pol.

Finally, we investigate the overall gain achievable by combining the
previous techniques. Fig.~\ref{fig:AIR}(c) compares the benchmark
AIR with that obtained by combining the SS procedure of Fig.~\ref{fig:AIR}(a)
and the PPN detection of Fig.~\ref{fig:AIR}(b) (with 4SC), either
with EDC or combined with DBP. The additional gain provided by SS
is smaller in this case, since the optimization accounts only for
intrachannel NLI and does not include PPN and DBP. Finally, including
the per-subcarrier power optimization proposed in \cite{garcia2020mismatched1,garciagomez2021mismatched}
further improves the gain up to 1.36~bits/sym/pol.

\section{Conclusions}

Although nonlinearity mitigation appears to be an elusive target,
many strategies have been devised over time, each addressing a specific
aspect of the problem. By combining an optimized input distribution,
a PPN-aware detection strategy, and including DBP, a gain of 1.36~bits/sym/pol
in the peak AIR is achieved compared to a linearly optimized system,
pushing the ultimate limit a little further and keeping alive the
hope of finding a truly optimal strategy. For simplicity, the input
is optimized, under some practical constraints, to minimize intrachannel
NLI in the absence of any other mitigation strategy. Higher gains
are expected by a full optimization that accounts also for interchannel
NLI and for the actual combination of processing and detection.\begin{spacing}{0.97}\setlength\bibitemsep{1pt}

\printbibliography
\end{spacing}
\end{document}